\documentclass[aps,prc,superscriptaddress,twoside,twocolumn,nofootinbib,showpacs,floatfix]{revtex4-1}
\usepackage{amsmath,amssymb}
\usepackage{graphicx}
\usepackage{braket,bm}
\usepackage{subfigure}
\allowdisplaybreaks
\begin{document}


\title{Understanding the structure of $d^*(2380)$ in chiral quark model}

\author{F. Huang}
\affiliation{School of Physical Sciences, University of Chinese Academy of Sciences, Beijing 101408, China}

\author{P. N. Shen}
\affiliation{Institute of High Energy Physics, Chinese Academy of Sciences, Beijing 100049, China}
\affiliation{Theoretical Physics Center for Science Facilities, CAS, Beijing 100049, China}
\affiliation{College of Physics and Technology, Guangxi Normal University, Guilin 541004, China}

\author{Y. B. Dong}
\affiliation{Institute of High Energy Physics, Chinese Academy of Sciences, Beijing 100049, China}
\affiliation{Theoretical Physics Center for Science Facilities, CAS, Beijing 100049, China}

\author{Z. Y. Zhang}
\affiliation{Institute of High Energy Physics, Chinese Academy of Sciences, Beijing 100049, China}
\affiliation{Theoretical Physics Center for Science Facilities, CAS, Beijing 100049, China}

\date{\today --- \jobname}


\begin{abstract}
The structure and decay properties of $d^*$ have been detailedly investigated in both the chiral SU(3) quark model and the extended chiral SU(3) quark model that describe the energies of baryon ground states and the nucleon-nucleon (NN) scattering data satisfactorily. By performing a dynamical coupled-channels study of the system of $\Delta\Delta$ and hidden-color channel (CC) with quantum numbers $I(J^P)=0(3^+)$ in the framework of the resonating group method (RGM), we find that the $d^*$ has a mass of about $2.38-2.42$ GeV and a root-mean-square radius (RMS) of about $0.76-0.88$ fm. The channel wave function is extracted by a projection of the RGM wave function onto the physical basis, and the fraction of CC component in the $d^*$ is found to be about $66\%-68\%$, which indicates that the $d^*$ is a hexaquark-dominated exotic state. 
Based on this scenario the partial decay widths of $d^*\to d \pi^0 \pi^0$ and $d^*\to d \pi^+\pi^-$ are further explicitly evaluated and the total width is then obtained by use of the branching ratios extracted from the measured cross sections of other possible decay channels. Both the mass and the decay width of $d^*$ calculated in this work are compatible with the data ($M\approx 2380$ MeV, $\Gamma\approx 70$ MeV) reported by WASA-at-COSY Collaboration.
\end{abstract}

\pacs{14.20.Pt,     
          13.75.Cs,    
          12.39.Jh,     
          13.30.Eg      
          }

\maketitle


\section{Introduction}

In recent years, the WASA experiments at CELSIUS \cite{CW2009} and COSY \cite{WASA2011} have identified a isoscalar resonance structure with mass $M\approx 2380$ MeV and width $\Gamma\approx 70$ MeV in the total cross sections of the double-pion fusion reactions $pn\to d\pi^0\pi^0$ and $pn\to d\pi^+\pi^-$. The data of angular dependences of these reactions are consistent with a spin-parity assignment of $J^P=3^+$ to this resonance structure. Later this resonance has further been observed in the non-fusion reactions $pn\to pn\pi^0\pi^0$ and $pn\to pp\pi^-\pi^0$ \cite{WASA2013,WASA2014-3}. In addition, the same resonance structure has also been reported in the double-pionic fusion reactions to the helium isotopes $pd\to {^3 \rm He}\,\pi^0\pi^0$, $pd\to {^3 \rm He}\,\pi^+\pi^-$, $dd\to {^4 \rm He}\,\pi^0\pi^0$ and $dd\to {^4 \rm He}\,\pi^+\pi^-$ \cite{WASA2015,WASA2006,WASA2012,WASA2009}. In the partial wave analysis of the neutron-proton ($np$) scattering, by incorporating the newly observed analyzing power data of the polarized $\vec{n}p$ scattering reaction, a resonance pole at $(2380 \pm 10) - i \,(40\pm 5)$ MeV in the $^3D_3$-$^3G_3$ coupled-channel waves has been obtained by the WASA-at-COSY Collaboration and the SAID Data Analysis Group \cite{WASA2014,WASA2014-2}, and henceforth this resonance has been denoted by $d^*(2380)$ following the convention used for excited nucleon states.

Even it is about 80 MeV lower than the threshold of $\Delta\Delta$, the mass of $d^*(2380)$ is still much higher than the thresholds of the $\Delta N\pi$ and $NN\pi\pi$ channels. However, the decay width of $d^*(2380)$ is about only 70 MeV, which is much smaller than the width of $\Delta$, $\Gamma_\Delta \sim 115$ MeV. Therefore, the $d^*(2380)$ resonance must be a very interesting state involving new physical mechanisms and is obviously worth investigating.

Theoretically, the possibility of the existence of $\Delta\Delta$ dibaryon states was first proposed in 1964 by Dyson and Xuong from a simple group classification based on SU(6) symmetry with no dynamical effects being taken into account \cite{Dyson64}. Since then, a lot of theoretical work has been devoted to explore the possible $\Delta\Delta$ dibaryon, but very few of them predicted a mass consistent with that of the $d^*(2380)$. The one that should be particularly mentioned is our work done in 1999 and later \cite{YZYS99}. There we have employed a constituent quark model with several sets of parameters
to perform dynamical calculations of the $\Delta\Delta$ system coupled with the hidden color channel (CC) in terms of the resonating group method (RGM), and a $\Delta\Delta$-CC bound state was predicted with a binding energy of about 40-80 MeV, which is compatible with the mass of $d^*(2380)$.

After the recent experimental observations, the $d^*(2380)$ has been investigated in several theoretical approaches \cite{Gal2013,Gal2014,HPW2014,CHX2015}. In Refs.~\cite{Gal2013,Gal2014}, Gal and Garcilazo carried out a $\pi N\Delta $ three-body-calculation and found the ${\mathcal
{D}_{03}}$ state as a dynamically generated pole at the right mass and with about 100 MeV for the width. In Ref.~\cite{HPW2014}, Huang, Ping and Wang performed a coupled-channel calculation in a chiral SU(2) quark model, and obtained an energy close to the observed value and a width of about 150 MeV which seems too large compared with the data.
 
In this work, we perform a further detailed investigation of the $\Delta\Delta-$CC system with quantum numbers $I(J^P)=0(3^+)$ within the framework of the RGM in both the chiral SU(3) quark model and the extended chiral SU(3) quark model. No free parameters are introduced here, as all the parameters needed are taken from our previous work \cite{DZYW2003,HZY2006}, which describes the energies of baryon ground states, the binding energy of deuteron, and the nucleon-nucleon (NN) scattering phases shifts quite well. Besides the binding energy of $d^*$, we concentrate on a detailed study of its structure and decay properties. After solving the dynamical coupled-channel RGM equation of $\Delta\Delta-$CC, the relative wave functions of $\Delta\Delta$ and CC channels are for the first time extracted by a projection of the RGM wave function onto the corresponding physical basis. The partial decay width of $d^*\to d \pi^0 \pi^0$ and $d^*\to d \pi^+\pi^-$ are then explicitly evaluated, and accordingly the total decay width of $d^*$ is obtained by use of the branching ratios extracted from the measured cross sections of other possible decay channels. The dependence of our theoretical results against the types of the confinement potential chosen in the model is also carefully tested.   

The paper is organized in the following. In the next section, we give a brief introduction of our chiral quark model and the determination of the model parameters. Then our results for the binding energy and structure of the $d^*$ are shown in Sec.~\ref{sec:binding}, together with some discussions. The partial and total decay width of $d^*$ are calculated and shown in Sec.~\ref{sec:decay}. Finally, a brief summary is given in Sec.~\ref{sec:summary}.


\section{Model and parameters}  \label{sec:model}

In Ref.~\cite{ZYS97}, the chiral SU(3) quark model was constructed by a linearly generalization of the linear SU(2) $\sigma$ model to the SU(3) case, there the quark and chiral fields interacting Lagrangian is written as
\begin{equation} \label{eq:Lagrangian}
{\cal L}_{I}^{\rm ch} = -g_{\rm ch} \bar{\psi} \left( \sum^{8}_{a=0} \lambda_a \sigma_a
 + i \gamma_5 \sum^{8}_{a=0} \lambda_a \pi_a \right) \psi,  
\end{equation}
with $g_{\rm ch}$ being the coupling constant of the quark and chiral fields interaction, $\psi$ the quark field, $\lambda_0$ the unitary matrix and $\lambda_{a}$ ($a=1,\cdots,8$) the generators of the flavor SU(3) group, $\sigma_{a}$ and $\pi_{a}$ ($a=0,1,\cdots,8$) the scalar and pseudo-scalar nonet fields, respectively. Clearly,
${\cal L}_{I}^{\rm ch}$ is invariant under the infinitesimal chiral SU(3)$_{L} \times$ SU(3)$_{R}$ transformation. As a consequence, the constituent quarks gain their constituent masses through the spontaneous chiral symmetry breaking and the Goldstone Bosons obtain their physical masses through the apparent chiral symmetry breaking caused by the non-zero masses of the current quarks. According to Eq.~(\ref{eq:Lagrangian}), the interactive Hamiltonian can be written as
\begin{eqnarray}  \label{eq:hamil}
H_{I}^{\rm ch} = g_{\rm ch} F(\bm q^{2}) \bar{\psi} \left( \sum^{8}_{a=0}
\lambda_a \sigma_a  + i \gamma_5 \sum^{8}_{a=0} \lambda_a \pi_a   \right) \psi,
\end{eqnarray}
where a form factor $F(\bm q^{2})$ is inserted to describe the chiral-field structure, and as usual it is taken as
\begin{eqnarray}\label{faca}
F(\bm q^{2})=\left(\frac{\Lambda^2}{\Lambda^2+\bm
q^2}\right)^{1/2},
\end{eqnarray}
with the cutoff mass $\Lambda$ indicating the chiral symmetry breaking scale \cite{amk91,abu91,emh91}.

From Eq.~(\ref{eq:hamil}), the quark-quark interaction potential induced by scalar and pseudo-scalar meson exchanges, $V^{\sigma_a}$ and $V^{\pi_a}$, can be directly derived. They mainly describe the non-perturbative QCD effect in the low-momentum medium-distance range. To study the hadron structure and hadron-hadron dynamics, an effective one-gluon-exchange (OGE) interaction $V^{\rm OGE}$ is still needed to govern the short-range perturbative QCD behavior, and a confinement potential $V^{\rm conf}$ is also needed to provide the non-perturbative QCD effect in the long distance. Finally the total Hamiltonian for a system with 6 quarks in the chiral SU(3) quark model can be written as
\begin{equation} \label{eq:hamiltonian}
H = \sum_{i=1}^6 T_i - T_G + \sum_{j>i=1}^6 \left(V^{\rm OGE}_{ij} + V^{\rm conf}_{ij} + V^{\rm ch}_{ij}\right),
\end{equation}
with $T_i$ being the kinetic energy operator for the $i$-th quark, $T_G$ the kinetic energy operator for the center of mass motion of the whole 6 quark system, and $V^{\rm ch}_{ij}$ the chiral fields induced effective interaction between the $i$-th quark and the $j$-th quark,
\begin{equation}
V_{ij}^{\rm ch} \,=\, \sum_{a=0}^8 V_{ij}^{\sigma_a} + \sum_{a=0}^8 V_{ij}^{\pi_a}.
\end{equation}

To study the short-range interaction mechanism, in Ref.~\cite{DZYW2003} the chiral SU(3) quark model has been extended to include the coupling between the quark and the vector meson fields as depicted by the following Lagrangian,
\begin{equation}
{\cal L}_I^{\rm chv} = -g_{\rm chv} \bar{\psi}\gamma_\mu \lambda_a
\rho^\mu_a \psi -\frac{f_{\rm chv}}{2M_N} \bar{\psi} \sigma_{\mu\nu}
\lambda_a
\partial^\mu \rho^\nu_a \psi.
\end{equation}
Here $\rho_{a}$ ($a=0,1,\cdots,8$) represent the vector nonet fields, and $g_{\rm chv}$ and $f_{\rm chv}$ are constants for vector and tensor coupling between quark and vector fields, respectively. This model is called the extended chiral SU(3) quark model, where the chiral fields induced effective interaction between the $i$-th quark and the $j$-th quark, $V^{\rm ch}$, now reads
\begin{equation}
V_{ij}^{\rm ch} \,=\, \sum_{a=0}^8 V_{ij}^{\sigma_a} + \sum_{a=0}^8 V_{ij}^{\pi_a} + \sum_{a=0}^8 V_{ij}^{\rho_a},
\end{equation}
with $V^{\rho_a}$ being the quark-quark interaction potential induced by vector-meson exchanges. Note that with the inclusion of vector-meson exchanges in the extended chiral SU(3) quark model, the contribution of OGE will be reduced automatically by fitting the mass splitting of N-$\Delta$.

The explicit expressions of the potentials $V^{\sigma_a}_{ij}$, $V^{\pi_a}_{ij}$, $V^{\rho_a}_{ij}$, $V^{\rm OGE}_{ij}$ and $V^{\rm conf}_{ij}$ for light quark systems are listed below:
\begin{eqnarray} \label{eq:V_sigma}
V^{\sigma_a}_{ij} = - C(g_{\rm ch},m_{\sigma_a},\Lambda) \left( \lambda_a^i\lambda_a^j \right) Y_1(m_{\sigma_a},\Lambda,r_{ij}),
\end{eqnarray}
\begin{eqnarray}
V^{\pi_a}_{ij}= \; && C(g_{\rm ch},m_{\pi_a},\Lambda)  \left(\lambda_a^i\lambda_a^j\right) \Big[Y_3(m_{\pi_a},\Lambda,r_{ij}) \nonumber \\
&& \times \, ({\bm \sigma}_i\cdot{\bm \sigma}_j) + H_3(m_{\pi_a},\Lambda,r_{ij}) S_{ij} \Big] \frac{m^2_{\pi_a}}{12m^2_q},
\end{eqnarray}
\begin{eqnarray} \label{eq:V_rho}
V^{\rho_a}_{ij} = \; && C(g_{\rm chv},m_{\rho_a},\Lambda) \left(\lambda_a^i\lambda_a^j\right) \Bigg\{
Y_1(m_{\rho_a},\Lambda,r_{ij}) \nonumber \\
&&  + \, \frac{m^2_{\rho_a}}{6m^2_q}  \Bigg[ \left(1+\frac{f_{\rm chv}}{g_{\rm chv}}\frac{2m_q}{M_N} + \frac{f^2_{\rm chv}}{g^2_{\rm chv}} \frac{m^2_q}{M^2_N} \right)  \nonumber \\
&&  \times \,  \bigg( Y_3(m_{\rho_a},\Lambda,r_{ij}) \left({\bm \sigma}_i\cdot{\bm \sigma}_j\right) -\frac{1}{2}  S_{ij}  \nonumber \\
&&  \times \, H_3(m_{\rho_a},\Lambda,r_{ij})  \bigg) \Bigg]\Bigg\},
\end{eqnarray}
\begin{eqnarray} \label{eq:V_OGE}
V^{\rm OGE}_{ij} = \; && \frac{1}{4}g_{i}g_{j}\left({\bm \lambda}^c_i\cdot{\bm \lambda}^c_j\right)
\Bigg\{ \Bigg[ \frac{1}{r_{ij}}-\frac{\pi}{m^2_q} \delta({\bm r}_{ij})  \nonumber \\
&& \times \, \left(1 + \frac{2}{3} {\bm \sigma}_i \cdot {\bm \sigma}_j  \right) - \frac{1}{4m^2_q} \frac{1}{r^3_{ij}} S_{ij} \Bigg] \Bigg\},
\end{eqnarray}
\begin{equation}
V_{ij}^{\rm conf}=-\left({\bm\lambda}_{i}^{c}\cdot{\bm \lambda}_{j}^{c}\right) \left(a_{ij}^{c}r_{ij}^2
-a_{ij}^{c0}\right),
\end{equation}
with $m_q$ being the quark mass, and
\begin{equation}
C(g_{\rm ch},m,\Lambda)=\frac{g^2_{\rm ch}}{4\pi} \frac{\Lambda^2}{\Lambda^2-m^2} m,
\end{equation}
\begin{equation} \label{x1mlr} 
Y_1(m,\Lambda,r)=Y(mr)-\frac{\Lambda}{m} Y(\Lambda r),
\end{equation}
\begin{equation}
Y_3(m,\Lambda,r)=Y(mr)-\left(\frac{\Lambda}{m}\right)^3 Y(\Lambda r),
\end{equation}
\begin{equation}
H_3(m,\Lambda,r)=H(mr)-\left(\frac{\Lambda}{m}\right)^3 H(\Lambda r),
\end{equation}
\begin{equation}
Y(x)=\frac{1}{x}e^{-x},
\end{equation}
\begin{equation}
H(x)=\left(1+\frac{3}{x}+\frac{3}{x^2}\right)Y(x),
\end{equation}
\begin{eqnarray}
S_{ij} = 3 \,{\bm\sigma}_i\cdot\hat{{\bm
r}}_{ij}\,{\bm\sigma}_j\cdot\hat{{\bm
r}}_{ij} - {\bm\sigma}_i\cdot{\bm\sigma}_j.
\end{eqnarray}
Note that in Eqs.~(\ref{eq:V_sigma}), (\ref{eq:V_rho}) and (\ref{eq:V_OGE}), the spin-orbit interactions are not included since they don't contribute in the calculation of the present work.

\begin{table}[tb]
\caption{\label{tab:para} Model parameters. The meson masses and the
cutoff masses: $m_{\sigma'}=980$ MeV, $m_{\epsilon}=980$ MeV, $m_{\pi}=138$ MeV, $m_{\eta}=549$ MeV, $m_{\eta'}=957$ MeV, $m_{\rho}=770$ MeV, $m_{\omega}=782$ MeV, and $\Lambda=1100$ MeV.}
\begin{tabular*}{\columnwidth}{@{\extracolsep\fill}lrrr}
\hline\hline
  & Ch. SU(3) & \multicolumn{2}{c}{Ext. Ch. SU(3)}  \\ \cline{3-4}
  &  & f/g=0 & f/g=2/3 \\
\hline
 $b_u$ (fm)  & 0.5 & 0.45 & 0.45 \\
 $m_u$ (MeV) & 313 & 313 & 313 \\
 $g_u^2$     & 0.766 & 0.056 & 0.132 \\
 $g_{\rm ch}$    & 2.621 & 2.621 & 2.621  \\
 $g_{\rm chv}$   &       & 2.351 & 1.973  \\
 $m_\sigma$ (MeV) & 595 & 535 & 547 \\
 $a^c_{uu}$ (MeV/fm$^2$) & 46.6 & 44.5 & 39.1 \\
 $a^{c0}_{uu}$ (MeV)  & $-$42.4 & $-$72.3 & $-$62.9 \\
\hline\hline
\end{tabular*}
\end{table}

The model parameters relevant to non-strange multi-quark systems are determined in the following way. The harmonic-oscillator width parameter $b_u$ in the Gaussian wave function for each $u$ or $d$ quark is taken to be $b_u=0.5$ fm in the chiral SU(3) quark model and $b_u=0.45$ fm in the extended chiral SU(3) quark model. The $u$ or $d$ quark mass is taken to be $m_{u(d)}=313$ MeV as usual. The coupling constant of the coupling between the quark field and the scalar and pseudo-scalar chiral fields, $g_{\rm ch}$, is determined according to the relation
\begin{equation}
\frac{g^{2}_{\rm ch}}{4\pi} = \left( \frac{3}{5} \right)^{2}
\frac{g^{2}_{NN\pi}}{4\pi} \frac{m^{2}_{u}}{M^{2}_{N}},
\end{equation}
with the empirical value $g^{2}_{NN\pi}/4\pi=13.67$. The coupling constant of the coupling between the quark field and the vector fields $g_{\rm chv}$ and the ratio of tensor coupling to vector coupling of vector-meson exchanges $f_{\rm chv}/g_{\rm chv}$ are taken 2 sets of values as done in our previous NN study \cite{DZYW2003}: $g_{\rm chv}=2.351$, $f_{\rm chv}/g_{\rm chv}=0$, and $g_{\rm chv}=1.973$, $f_{\rm chv}/g_{\rm chv}=2/3$. The masses of all the mesons are taken to be the experimental values, except for the $\sigma$ meson, whose mass is
fixed by fitting the NN scattering data. The cutoff radius $\Lambda^{-1}$ is taken to be the value close to the chiral symmetry breaking scale \cite{amk91,abu91,emh91}. After the parameters of chiral fields are fixed, the coupling constant $g_u$ of OGE is then determined by the mass split of $N$-$\Delta$. The confinement strength $a^{c}_{uu}$ and the zero-point energies $a^{c0}_{uu}$ are fixed by the stability condition and the mass of nucleon,  respectively.


All the model parameters relevant to the present work are tabulated in Table~\ref{tab:para}, where the first set is for the original chiral SU(3) quark model, the second and third sets are for the extended chiral SU(3) quark model. One sees that in the extended chiral SU(3) quark model, due to the inclusion of vector-meson exchanges, the coupling constant of OGE is rather small compared with that in the original chiral SU(3) quark model. 

We emphasize that the values of all the model parameters listed in Table~\ref{tab:para} are taken from our previous work \cite{DZYW2003,HZY2006}, and they can give a satisfactory description of the energies of the octet and decuplet baryon ground states, the nucleon-nucleon scattering phase shifts in the low energy region ($\sqrt{s} \leqslant 2m_N+200$ MeV) and the binding energy of the deuteron. No additional parameter is introduced in the present work, which we think would help to improve the reliability of the calculated results.

The RGM \cite{HZY2004}, a standard method suitable for the study of the interaction between complex particles or clusters, will be employed to investigate the interaction property of the system of $\Delta\Delta$ and CC. Here $\Delta$ and C are described in quark model as states with the following symmetry and quantum numbers
\begin{align*}
\Delta: & \quad \left(0s\right)^3 [3]_{\rm orb}, S=3/2, I=3/2, C=(00),  \\[5pt]
C: & \quad \left(0s\right)^3 [3]_{\rm orb}, S=3/2, I=1/2, C=(11),
\end{align*}
where $\left(0s\right)$ is the harmonic-oscillator shell wave function,  $[3]_{\rm orb}$ represents the symmetry in orbit space for each cluster, and $S$, $I$ and $C$ are quantum numbers of spin, isospin and color for each cluster. In the framework of RGM, the trial microscopic wave function of the whole six-quark system of $\Delta\Delta$-CC with spin $S=3$ and isospin $T=0$ can be written as
\begin{multline}
\Psi_{6q} =  {\cal A} \left[  \phi_\Delta\!\left( {\bm \xi}_1, {\bm \xi}_2 \right)
\phi_\Delta\!\left( {\bm \xi}_4, {\bm \xi}_5 \right) \eta_{\Delta\Delta}\!
\left({\bm r}\right) + \right. \\[6pt]
 \left. \phi_{\rm C}\!\left( {\bm \xi}_1, {\bm \xi}_2 \right)
\phi_{\rm C}\!\left( {\bm \xi}_4, {\bm \xi}_5 \right) \eta_{\rm CC}\!
\left({\bm r}\right)  \right]_{S=3, I=0, C=\left(00\right)}. \label{Eq:wavfun}
\end{multline}
Here ${\cal A}$ is the antisymmetrizer required by the Pauli exclusion principle, $\phi_{\Delta{\rm (C)}}$ 
the antisymmetrized internal wave functions of the (123) ((456))
3-quark cluster with ${\bm \xi}_i$ ($i=1,2 ~(4,5)$) being its internal
Jacobi coordinates, and $\eta_{\Delta\Delta {\rm (CC)}}$ the coordinate wave function of the relative motion of two clusters $\Delta\Delta$(CC) which is determined completely by the interacting dynamics of the whole six-quark system.

Expanding the unknown $\eta_{\Delta\Delta {\rm (CC)}}$ in Eq.~(\ref{Eq:wavfun}) by employing well-defined basis wave functions, such as Gaussian functions, one can dynamically solve the RGM equation for a bound state problem,
\begin{equation}
\braket{\delta\Psi_{6q} | H-E | \Psi_{6q}} = 0, \label{eq:RGM-bound}
\end{equation}
to get the binding energy $E$ and the corresponding RGM 6-quark wave function $\Psi_{6q}$ for the $\Delta\Delta$-CC system.

For more details about our chiral quark model and the RGM, we refer the readers to our previous work \cite{ZYS97,DZYW2003,HZY2004,HZ2004}.

\section{Energy and structure of $d^*$}  \label{sec:binding}

\begin{table*}[tb] 
\caption{\label{tab:dstarbe} Binding energy, root-mean-square radius (RMS) of 6 quarks, and fraction of channel wave function for $d^*$ in the chiral SU(3) quark model and extended chiral SU(3) quark model with ratios of tensor coupling to vector coupling f/g=0 and f/g=2/3 for vector meson fields.}
\begin{tabular*}{\textwidth}{@{\extracolsep\fill}lrrrrrr} 
 \hline\hline
   & \multicolumn{3}{c}{$\Delta\Delta$ ($L=0,2$) } & \multicolumn{3}{c}{$\Delta\Delta\,-\,$CC  ($L=0,2$) } \\[1pt] \cline{2-4} \cline{5-7}
   & SU(3) & Ext. SU(3) & Ext. SU(3) & SU(3) & Ext. SU(3) & Ext. SU(3) \\
   &            &   (f/g=0)  &  (f/g=2/3)  &    &   (f/g=0)  &  (f/g=2/3)   \\ \hline
 Binding energy (MeV) &  28.96  &  62.28  &  47.90 & 47.27 & 83.95 & 70.25 \\ 
 RMS of $6q$ (fm) &   0.96   &  0.80  &  0.84  & 0.88 & 0.76 & 0.78 \\ 
 Fraction of ($\Delta\Delta$)$_{L=0}$ ($\%$) & 97.18 &  98.01  & 97.71 &  33.11 & 31.22 & 32.51 \\
 Fraction of ($\Delta\Delta$)$_{L=2}$ ($\%$) & 2.82  & 1.99  &  2.29 & 0.62 & 0.45 &  0.51 \\ 
 Fraction of (CC)$_{L=0}$ ($\%$) &   &    &   &  66.25 & 68.33 & 66.98 \\
 Fraction of (CC)$_{L=2}$ ($\%$) &   &    &   &  0.02 & 0.00 & 0.00 \\  
 \hline\hline
\end{tabular*}
\end{table*}

\begin{table}[tb]
\caption{\label{tab:dbe} Binding energy, root-mean-square radius (RMS) of 6 quarks, and fraction of channel wave function for deuterons in the chiral SU(3) quark model and extended chiral SU(3) quark model with ratios of tensor coupling to vector coupling f/g=0 and f/g=2/3 for vector meson fields. }
\begin{tabular*}{\columnwidth}{@{\extracolsep\fill}lrrr}
 \hline\hline
    &  SU(3) & \multicolumn{2}{c}{Ext. SU(3)}  \\ \cline{3-4}
    &            &  (f/g=0)       & (f/g=2/3)     \\ \hline
 Binding energy (MeV) &  2.09  &  2.24  &  2.20 \\ 
 RMS of $6q$ (fm) &    1.38   &  1.34  &   1.35 \\ 
 Fraction of (NN)$_{L=0}$ ($\%$) & 93.68 &  94.66  & 94.71 \\
 Fraction of (NN)$_{L=2}$ ($\%$) & 6.32  &  5.34  &  5.29 \\  \hline\hline
\end{tabular*}
\end{table}

To get the binding behavior of $\Delta\Delta$-CC system with $I(J^P)=0(3^+)$, we dynamically solve Eq.~(\ref{eq:RGM-bound}), the RGM equation for a bound state problem, in both the chiral SU(3) quark model and the extended chiral SU(3) quark model with the parameters listed in Table~\ref{tab:para}. The results are tabulated in Table~\ref{tab:dstarbe} \cite{Huang2014}. Let's first focus on the results from a $\Delta\Delta$ ($L=0,2$) double-channel calculation. One sees that the $\Delta\Delta$ with $I(J^P)=0(3^+)$ is indeed bound with a binding energy of about $30-60$ MeV. The corresponding root-mean-square radius (RMS) of this bound state is found to be around $0.80-0.96$ fm. Now Let's concentrate on the results from a $\Delta\Delta$-CC ($L=0,2$) quadruple-channel calculation. One sees that the coupling to the CC channel further results in an increment of about $20$ MeV to the binding energy of the bound state of this system, called $d^*$, and a considerable decrement of the RMS of this state. Finally, the mass of this bound state reaches $2.38-2.42$ GeV and the RMS shrinks to $0.76-0.88$ fm. This clearly shows that $d^*$ is a $\Delta\Delta$-CC deeply bound and compact state where the coupling to the CC channel plays a significant role. For comparison, we also list our results for deuteron obtained by using the same sets of parameters in Table~\ref{tab:dbe}. One sees that the deuteron has a binding energy of about $2.09-2.24$ MeV and a RMS of about $1.34-1.38$ fm, both agree well with the experimental data. Clearly, unlike the $\Delta\Delta$-CC system with $I(J^P)=0(3^+)$, the deuteron is weekly bound and loose.

Our further analysis shows that the distinctive features of the $\Delta\Delta$ system with $I(J^P)=0(3^+)$, i.e. it is deeply bound and couples strongly to CC, are due to both the quark exchange effect and the short-range interaction being ``attractive" in this system. 

As the internal wave functions $\phi_{\Delta(\rm C)}$ are antisymmetrized already for each cluster, the antisymmetrizer in Eq.~(\ref{Eq:wavfun}) for 6 identical quarks is usually simplified as
\begin{equation}
{\cal A} = 1 - 9 P_{36}, \label{eq:anti-A}
\end{equation}
and its averaged value in spin-flavor-color space, $\braket{{\cal A}^{sfc}}$, is usually used to characterize the quark-exchange effect for a system composed of composite clusters \cite{LSZY2001}. In general, the value of $\braket{{\cal A}^{sfc}}$ is highly dependent on the quantum numbers of a considered complex system, and is in the range
\begin{equation}
\braket{{\cal A}^{sfc}} \in \left[0,2\right].
\end{equation}
$\braket{{\cal A}^{sfc}} \sim 1$ means that there is almost no quark-exchange effect between two clusters, while $\braket{{\cal A}^{sfc}} \sim 2$ means that the quark-exchange effect is highly strong and helpful to drag two clusters together to form a compact state. For the NN $^3S_1$ partial wave, one gets
\begin{equation}
\braket{NN | {\cal A}^{sfc} | NN}_{S=1,I=0} = 10/9 \sim 1, \label{eq:quark-d}
\end{equation}
i.e. the quark-exchange effect is almost negligible in the deuteron.
For the $\Delta\Delta$ system with $I(J^P)=0(3^+)$, one has
\begin{equation}
\braket{\Delta\Delta | {\cal A}^{sfc} | \Delta\Delta}_{S=3,I=0} = 2, \label{eq:quark-dstar}
\end{equation}
i.e. such a state is very special as it has the strong quark-exchange effect, and in particular, it tends to drag two $\Delta$'s together. It is thus natural to speculate that such an ``attraction" could result in a $\Delta\Delta$ bound state. Note that such an analysis is model independent and purely based on the symmetry property of the system.

Apart from the strong and ``attractive" quark-exchange effect, the property of the short-range interaction is another characteristic which makes the $\Delta\Delta$ with $I(J^P)=0(3^+)$ rather special. In 1980 Oka and Yazaki claimed that in all the non-strange two-baryon systems, the $\Delta\Delta$ system with $I(J^P)=0(3^+)$ is the only one in which the OGE provides strong attraction at short range \cite{OY80}. In our chiral SU(3) quark model, we find that the OGE indeed provides short-range attraction. In the extended chiral SU(3) quark model, although the OGE is largely reduced, the short-range interaction is still attractive and the attraction is even much stronger, as the vector meson exchanges (VMEs) are also strongly attractive at short range. 

Considering the facts that the quark exchange effect is strong and ``attractive" and the short-range interaction provided by OGE or VMEs is also attractive, the $\Delta\Delta$ system with $I(J^P)=0(3^+)$ is indeed a highly unusual system. It couples to the CC channel strongly as the two interacting $\Delta$s could be dragged close enough by strong attraction. 


The NN $^3S_1$ partial wave has a rather different feature than the $\Delta\Delta$ with $I(J^P)=0(3^+)$. We know from Eq.~(\ref{eq:quark-d}) that in NN $^3S_1$ partial wave, the quark-exchange effect is very weak (almost negligible). Moreover, the short-range interactions stemming from OGE and VMEs for this partial wave are all repulsive. Therefore, the NN $^3S_1$ partial wave can only get attraction in the medium-  and long-range through $\sigma$ and $\pi$ meson exchanges, and as a result, the deuteron is loosely bound and hardly couples to the CC channel. 


Although the theoretical mass of $d^*$ from our model is already comparable with the data, i.e. $M \approx 2.38$ GeV as reported by WASA-at-COSY Collaboration, its unusual narrow width, $\Gamma \approx 70$ MeV, still needs to be understood. 
Our $\Delta\Delta$-CC ($L=0,2$) coupled-channel RGM calculation shows that the coupling to the CC channel will result in an increment of about 20 MeV to the binding energy of $\Delta\Delta$. This implies that the configuration of CC might be substantial in $d^*$. 
For a concrete investigation of the decay width of $d^*$ and a thorough understanding of its internal structure, extracting the relative wave functions of both the $\Delta\Delta$ and the CC channels becomes essential.

\begin{figure*}[tb]
\centering
\subfigure[~$d^*$]{
\label{fig:subfig:b} 
\includegraphics[width=0.49\textwidth]{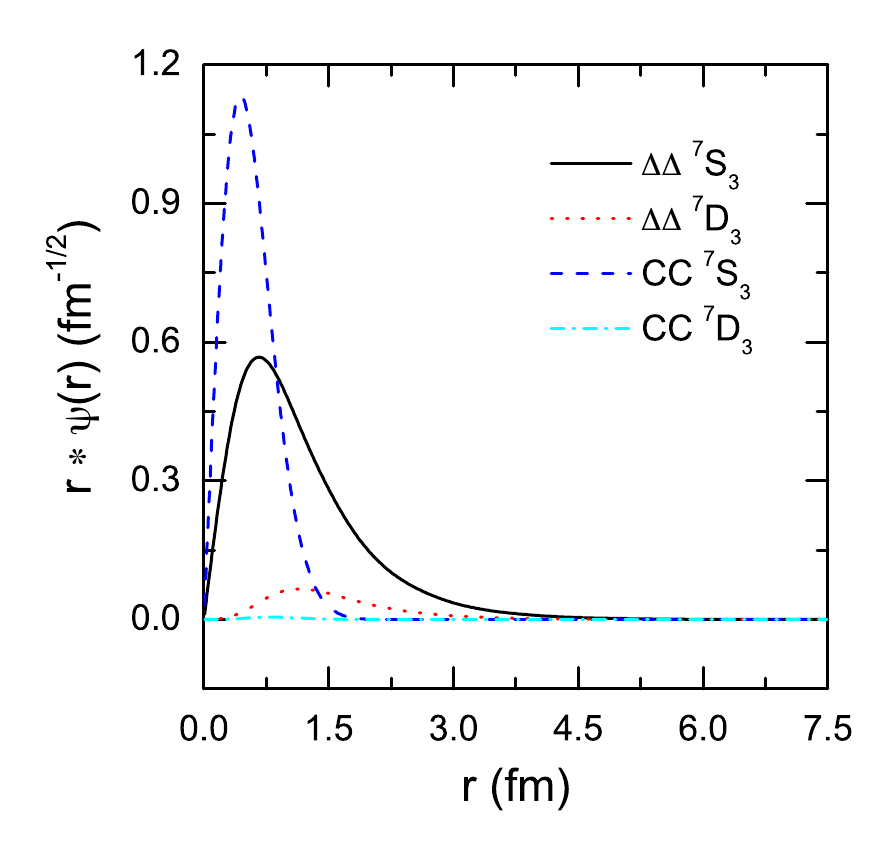}}
\subfigure[~$d$ (deuteron)]{
\label{fig:subfig:a} 
\includegraphics[width=0.49\textwidth]{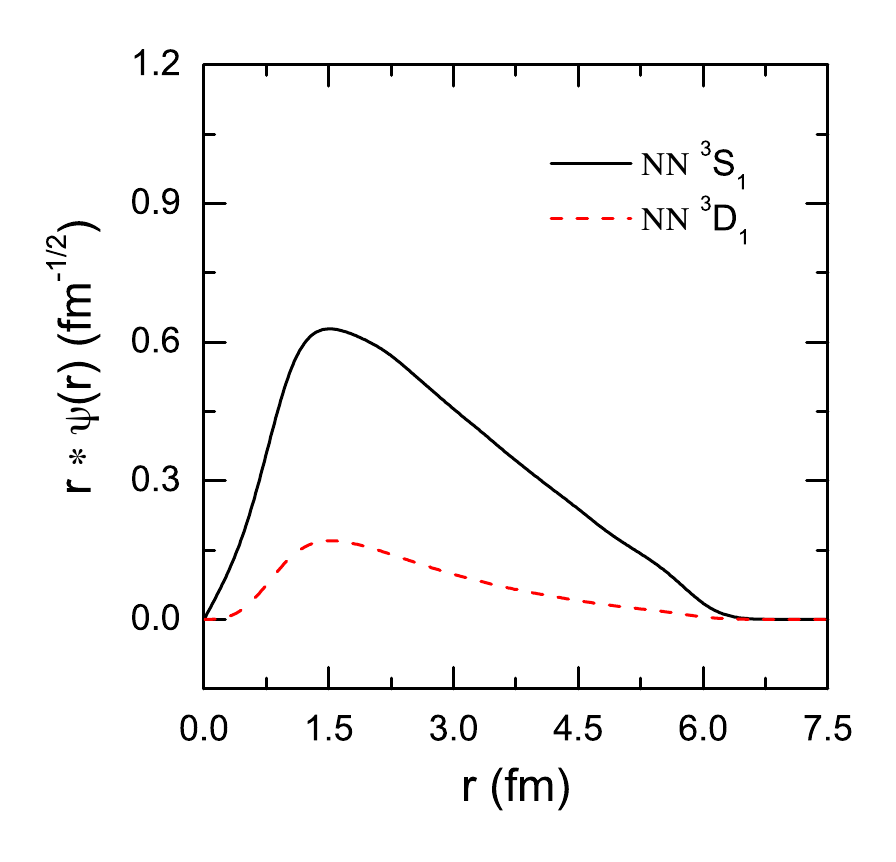}}
\caption{Relative wave functions in the extended chiral SU(3) quark model with f/g=0 for the $d^*$ (left) and deuteron (right).}
\label{fig:wf} 
\end{figure*}

Inserting Eq.~(\ref{eq:anti-A}) into Eq.~(\ref{Eq:wavfun}), one writes the RGM wave functions for $\Delta\Delta$ and CC channels as
\begin{multline}
\Psi_{6q} =  \left(1-9P_{36}\right) \left[  \phi_\Delta\!\left( {\bm \xi}_1, {\bm \xi}_2 \right)
\phi_\Delta\!\left( {\bm \xi}_4, {\bm \xi}_5 \right) \eta_{\Delta\Delta}\!
\left({\bm r}\right) + \right. \\[6pt]
 \left. \phi_{\rm C}\!\left( {\bm \xi}_1, {\bm \xi}_2 \right)
\phi_{\rm C}\!\left( {\bm \xi}_4, {\bm \xi}_5 \right) \eta_{\rm CC}\!
\left({\bm r}\right)  \right]_{S=3, I=0, C=\left(00\right)}. \label{Eq:wavfun2}
\end{multline}
Clearly that the above wave functions for $\Delta\Delta$ and CC are not orthogonal to each other due to the transition between these two channels caused by quark exchanges, and thus they are not suitable to be used directly to clarify the components of $\Delta\Delta$ and CC in $d^*$. Instead, the relative wave function in physics basis, namely the channel wave function in the quark cluster model, is introduced \cite{Kusainov91,Glozman93,Stancu97}. In the present case, the channel wave functions for $\Delta\Delta$ and CC are defined as
\begin{subequations}
\begin{align}
\chi_{\Delta\Delta}\!\left({\bm r}\right) & \equiv \Braket{ \phi_\Delta\!
\left( {\bm \xi}_1, {\bm \xi}_2 \right) \phi_\Delta\!\left( {\bm \xi}_4,
{\bm \xi}_5 \right) | \Psi_{6q} },  \label{Eq:6q-1}  \\[6pt]
\chi_{\rm CC}\!\left({\bm r}\right) & \equiv \Braket{ \phi_{\rm C}\!
\left( {\bm \xi}_1, {\bm \xi}_2 \right) \phi_{\rm C}\!\left( {\bm
\xi}_4, {\bm \xi}_5 \right) | \Psi_{6q} }, \label{Eq:6q-2}
\end{align}
\end{subequations}
where $\phi_{\Delta({\rm C})}$ and $\Psi_{6q}$ are the antisymmetrized internal wave function for the cluster of $\Delta({\rm C})$ and the total RGM wave function for the whole six-quark system, respectively, as shown in Eq.~(\ref{Eq:wavfun}), and $\langle \dots \rangle$ denotes the integral over all the
internal coordinates of $3q$ clusters. Then the wave function of $d^*$
can be simply abbreviated and expanded as
\begin{align}
\Psi_{d^*} &= \Ket{\Delta\Delta} \chi_{\Delta\Delta}\!\left({\bm r}\right) +
\Ket{\rm CC} \chi_{\rm CC}\!\left({\bm r}\right) \nonumber \\[5pt]
&=\sum_{L=0,2} \left[\Ket{\Delta\Delta} \frac{\chi^L_{\Delta\Delta}\!
\left(r\right) }{r} + \Ket{\rm CC} \frac{\chi^L_{\rm CC}\!\left(r\right) }{r}
\right]  Y_{L0}\!\left({\hat {\bm r}}\right) .  \label{Eq:DD-CC}
\end{align}
From Eqs.~(\ref{Eq:wavfun}), (\ref{eq:anti-A}), (\ref{Eq:6q-1}) and (\ref{Eq:6q-2}),
one sees that $\chi^L_{\Delta\Delta}\!\left(r\right)$ contains
contributions from not only the $\Delta\Delta$ direct term and 
exchange term but also the CC exchange term (namely, the contribution due to
the transition from the CC channel to the $\Delta\Delta$ channel).
Similarly, $\chi^L_{\rm CC}\!\left(r\right)$ contains contribution
from not only the CC direct term and exchange term but also the $\Delta\Delta$ exchange
term. In other words, the antisymmetrization effect in
Eq.~(\ref{Eq:wavfun}) has already been absorbed into
$\chi^L_{\Delta\Delta}\!\left(r\right)$ and $\chi^L_{\rm
CC}\!\left(r\right)$. It is obvious that the $\Delta\Delta$ and CC
components in $d^*$ as depicted in Eq.~(\ref{Eq:DD-CC}) are
orthogonal to each other, and hence are suitable to be employed to
discuss the spatial distribution of $d^*$ and its individual components of $\Delta\Delta$ and CC. 

We mention that this projection approach might not be very rigorous, but to a certain extent it is still a quite reasonable and efficient way to discuss the internal structure or individual components of a state composed of composite clusters, as it absorbs the quark exchange effects into the channel wave function (relative wave function in physical basis) and hence largely reduces the complexity of the relevant calculation and discussion based on the complex RGM wave function of the whole $6q$ system.

The partial-wave projected relative wave functions $\chi^L_{\Delta\Delta}\!\left(r\right)$ and $\chi^L_{\rm CC}\!\left(r\right)$ ($L=0,2$) obtained in the extended chiral SU(3) quark model with f/g=0 are plotted in Fig.~\ref{fig:wf}, where for comparison the relative wave functions for deuteron calculated in the same way are also plotted. The results calculated by use of the other two sets of parameters as tabulated in Table~\ref{tab:para} are quite similar. From Fig.~\ref{fig:wf} one sees that $d^*$ is rather narrowly distributed and it has a maximal distribution located around $0.7$ fm for $\Delta\Delta$ ($L=0$) and $0.4$ fm for CC ($L=0$), respectively. The deuteron, on the other hand, is widely distributed with a maximal distribution located around $1.4$ fm. We also see that the deuteron has a considerable $D$-wave contribution which is actually the crucial point making the deuteron loosely bound, while for $d^*$ the $D$-wave contribution is much smaller.

The fractions of each individual channel in total wave functions can be extracted by an integral of the square of the normalized relative wave functions. The calculation is first done for deuteron and the results are listed in Table~\ref{tab:dbe}. One sees that the $S$-wave accounts for $\sim 95\%$ and the $D$-wave accounts for $\sim 5\%$ in the deuteron, which as we all know is quite reasonable. Similar calculation is then done for $d^*$ and the corresponding results are tabulated in Table~\ref{tab:dstarbe}. One sees that the CC component dominates the structure of the $d^*$, as the fraction of the CC channel in $d^*$ is about $66\%-68\%$. Note that according to symmetry, a pure hexaquark state of the $\Delta\Delta$-CC system with isospin $I=0$ and spin $S=3$ reads
\begin{equation}
[6]_{\rm orb} [33]_{IS=03} = \sqrt{\frac{1}{5}} \Ket{\Delta\Delta}_{IS=03} + \sqrt{\frac{4}{5}} \Ket{\rm CC}_{IS=03},
\end{equation}
which indicates that the fraction of CC channel in a pure hexaquark state is $80\%$. It is thus fair to say that $d^*$ is a hexaquark-dominated exotic state as it has a CC configuration of about $66\%-68\%$. This finding is of great interest. It provides a chance to naturally explain that $d^*$, although located above the thresholds of the $\Delta N \pi$ and $NN\pi\pi$ channels, has a relatively narrow width since its dominated CC components cannot be subject to a direct break-up decay. 
A detailed calculation of the partial decay widths of $d^*\to d\pi^+\pi^-$ and $d^*\to d\pi^0\pi^0$ \cite{Dong2015} will be present in the next section, where we will see that apart from the mass, the picture we proposed here for $d^*$ also results in a good agreement of the decay width compared with the data reported by WASA-at-COSY Collaboration.

\begin{table}[tb] 
\caption{\label{tab:confinement} Binding energy, root-mean-square radius (RMS) of 6 quarks, and fraction of channel wave function for $d^*$ in the extended chiral SU(3) quark model with ratio of tensor coupling to vector coupling f/g=0 and the types of confinement potential taken to be quadratic, linear and error functional, respectively.}
\begin{tabular*}{\columnwidth}{@{\extracolsep\fill}lrrr} 
 \hline\hline
   & \multicolumn{3}{c}{$\Delta\Delta\,-\,$CC  ($L=0,2$) } \\[1pt] \cline{2-4}
   &    $r^2$        &   $r$  &  Erf\,($r$)   \\ \hline
 Binding energy (MeV) &  83.95 & 86.17 & 87.63 \\ 
 RMS of $6q$ (fm) &   0.76 & 0.76 & 0.76 \\ 
 Fraction of ($\Delta\Delta$)$_{L=0}$ ($\%$) &   31.22 & 30.66 & 30.41 \\
 Fraction of ($\Delta\Delta$)$_{L=2}$ ($\%$) &  0.45 & 0.42 &  0.40 \\ 
 Fraction of (CC)$_{L=0}$ ($\%$) &   68.33 & 68.92 & 69.19 \\
 Fraction of (CC)$_{L=2}$ ($\%$) &  0.00 & 0.00 & 0.00 \\  
 \hline\hline
\end{tabular*}
\end{table}

A conjecture that $d^*$ should have an unconventional origin and that the CC configuration will suppress its decay width has also been mentioned by Bashkanov, Brodsky and Clement in Ref.~\cite{BBC2013}. Our microscopic calculation presented here supports their argument and moreover, our dynamical results show explicitly that $d^*$ has about $2/3$ hidden-color configurations and thus is a hexaquark-dominated exotic state. Physically, the peculiar structure of $d^*$ proposed here is not difficult to be understood if we keep in mind that the $\Delta\Delta$ system with $I(J^P)=0(3^+)$ is a highly unusual system as both the quark exchange effect and the short-range interaction of it are attractive, which makes this system strongly compact and promotes a strong coupling to the CC channel.
 
Although no free parameter is introduced in the present work, a particular type of confinement, i.e. a quadratic one, is used as done in our previous work \cite{ZYS97,DZYW2003,HZY2004,HZ2004}. As is well known that the results in the RGM calculation of a single-channel composed of two color-singlet clusters are independent of any types of the confinement potential, but here in $d^*$, as the CC channel is important, we think it should be checked how different types of confinement potential affect the results of $d^*$. Therefore, we perform two additional analysis by using a linear confinement and an error functional confinement in the extended chiral SU(3) quark model with ratio of tensor coupling to vector coupling $f_{\rm chv}/g_{\rm chv}=0$, and the results are shown in Table~\ref{tab:confinement} compared with those obtained by using a quadratic confinement. One sees that the energy of $d^*$ varies less than 4 MeV, while the RMS of 6 quarks and the fraction of each channel in $d^*$ almost keep unchanged when the quadratic confinement is replaced by a linear one or an error functional one. This clearly shows that our results of $d^*$ are rather stable against the types of the confinement potential, the reason for which is that the $\Delta\Delta$-CC system with $I(J^P)=0(3^+)$ is rather compact and thus is not sensitive to the long-range confinement potential.

\section{Decay width of $d^*$}  \label{sec:decay}

\begin{figure}[tb]
\vglue 0.6cm
\includegraphics[width=0.4\textwidth]{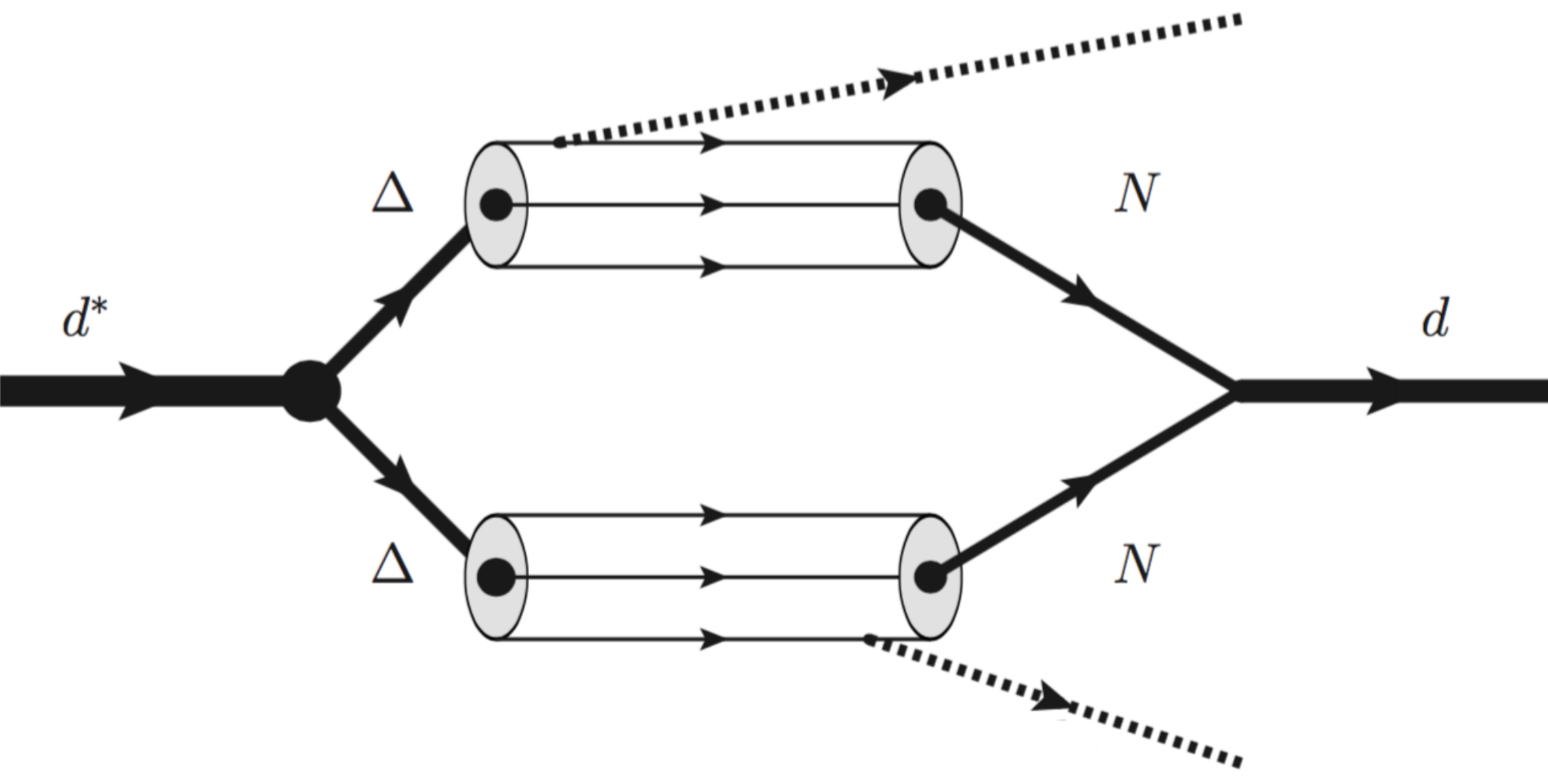}
\caption{Schematic diagram for the decay of $d^*\to d\pi\pi$. }  \label{fig:decay}
\end{figure}

In Sec.~\ref{sec:binding}, we have obtained the relative wave functions of $\Delta\Delta$ and CC channels in $d^*$. 
In this section, we perform an explicit calculation of the partial decay widths of $d^*\to d\pi^+\pi^-$ and $d^*\to d\pi^0\pi^0$ by use of these channel wave functions.

As we mentioned before, the CC component cannot result in a direct break-up decay.
In the lowest order, the decay process $d^*\to d\pi\pi$ occurs in such a way that each of the two $\Delta$'s in $d^*$ emits a pion and the remaining two nucleons form a deuteron.  
A schematic diagram of this process is depicted in Fig.~\ref{fig:decay}.

Following Refs.~ \cite{Itonaga1996,Obukhovsky1997}, the Hamiltonian density for the quark-quark-pion interaction in non-relativistic approximation reads
\begin{equation}
{\cal H}_{qq\pi} = g_{qq\pi} {\bm \sigma} \cdot {\bm k}  {\bm \tau} \cdot {\bm \pi} \, \frac{1}{\left(2\pi\right)^{3/2}\sqrt{2\omega}},
\end{equation}
where $g_{qq\pi}$ is the coupling constant to be fixed by the decay width of $\Delta\to N\pi$, and $\left(\omega, {\bm k}\right)$ is the four-momentum of the pion meson. This Hamiltonian density gives the $qq\pi$ vertex as shown in Fig.~\ref{fig:decay}. Further by using the relative wave functions for $\Delta\Delta$ in $d^*$ and NN in $d$, the transition amplitude ${\cal M}_{fi}$ of the decay process $d^*\to d\pi\pi$ can be obtained straightforwardly \cite{Dong2015}, and the partial decay width of this reaction can then be evaluated by
\begin{multline}
\Gamma_{d^*\to d\pi\pi} = \frac{1}{2!} \int
d{\bm k}_1 d{\bm k}_2 d{\bm p}_{d} \left(2\pi\right) \delta^3\!\left(\bm{k}_1+\bm{k}_2+\bm{p}_d\right)  \\[3pt]
\times \delta\!\left(\omega_{k_1}+\omega_{k_2}+E_d-M_{d^*}\right) \left |
\overline{{\cal M}_{fi}} \right | ^2,  \label{eq:gamma} 
\end{multline}
with $\left(\omega_i, {\bm k}_i\right)$ $(i=1,2)$ being the four-momentum for two pions, and $\left(E_d, {\bm p}_d\right)$ being the four-momentum for deuteron. 

By using the relative wave functions provided by our model as plotted in Fig.~\ref{fig:wf}, the partial decay widths for both $d^*\to d\pi^+\pi^-$ and $d^*\to d\pi^0\pi^0$ are obtained by a calculation of Eq.~(\ref{eq:gamma}) and the resultant numerical results are
\begin{align}
 & \Gamma_{d^*\to d\pi^+\pi^-}\approx 16.8~ {\rm MeV}, \\[5pt]
 & \Gamma_{d^*\to d\pi^0\pi^0}\approx 9.2~ {\rm MeV}.
 \end{align}
The partial widths for other decay processes can then be obtained by using the branching ratios extracted from the cross sections of those channels measured by experiments \cite{BCS2015,BCS2015-2}, and finally a sum of them results in a total decay width of $d^*$, $\Gamma \sim 69$ MeV, which is in consistent with the experimental value $\Gamma \approx 70$ MeV. 
This justifies the scenario we proposed for $d^*$, i.e. it has about 2/3 CC component and thus is a hexaquark-dominated exotic sate. 

Of course a more serious calculation for the partial decay widths for all the decay channels should be evaluated directly by using the relative wave functions of $d^*$ provided in the present work. Research along this line is ongoing.

\section{Summary}  \label{sec:summary}

In this work, we have detailedly investigated the structure and decay properties of $d^*$ in both the chiral SU(3) quark model and the extended chiral SU(3) quark model.

Based on a microscopic investigation of the $\Delta\Delta$-CC $I(J^P)=0(3^+)$ system with no additional parameters besides those already fixed in the study of NN scattering phase shifts,  it is found that the $d^*$ has a mass of about $2.38-2.42$ GeV and a root-mean-square radius of about $0.76-0.88$ fm. For the first time, the relative wave functions in physical basis for each individual channels in $d^*$ are extracted, and it is found that the $d^*$ has a CC fraction of about $66\%-68\%$, which manifests that the $d^*$ is a hexaquark-dominated exotic state. All these results are tested to be rather stable against the types of the confinement potential selected in the model.

Further based on the scenario we proposed for $d^*$, the partial decay widths of $d^*\to d \pi^0 \pi^0$ and $d^*\to d \pi^+\pi^-$ are further explicitly evaluated. A total decay width $\Gamma \sim 69$ MeV is then obtained by the use of the branching ratios extracted from the cross sections measured for the corresponding decay channels.

Both the theoretical mass and width of $d^*$ calculated in our model are in consistent with the data reported recently by the WASA-at-COSY Collaboration, which strongly supports the scenario we proposed for $d^*$, i.e. it has about 2/3 CC components and thus is a hexaquark-dominated exotic state. If this scenario can be further verified, the $d^*(2380)$ will be the first hexaquark-dominated exotic state one has ever found, and it may open a door to new physical phenomena. We look forward to more theoretical work from other approaches and more experimental work from other laboratories to further confirm the properties of the $d^*(2380)$.


\begin{acknowledgments}
We are grateful for constructive discussions with Profs. S. Brodsky, A. Buchmann, C. H. Chang, H. Clement, H. B. Li, C. P. Shen, R. L. Workman, and Q. Zhao. 
This work is partly supported by the National Natural Science Foundation of China under grants Nos. 11475181, 11475192, 11035006 and 11165005, the fund provided to the Sino-German CRC 110 “Symmetries and the Emergence of Structure in QCD” project by the DFG, and the IHEP Innovation Fund under the No. Y4545190Y2. F.H. is grateful to the support of the One Hundred Person Project of the University of Chinese Academy of Sciences.
\end{acknowledgments}




\begin{thebibliography}{99}
%
%
\bibitem{CW2009}
M. Bashkanov  {\it et al.}, Phys. Rev. Lett. {\bf 102}, 052301 (2009).
%
\bibitem{WASA2011}
P. Adlarson {\it et al.} (WASA-at-COSY Collaboration), Phys. Rev. Lett. {\bf 106}, 242302 (2011).
%
\bibitem{WASA2013}
P. Adlarson {\it et al.} (WASA-at-COSY Collaboration), Phys. Rev. C {\bf 88}, 055208 (2013).
%
\bibitem{WASA2014-3}
P. Adlarson {\it et al.} (WASA-at-COSY Collaboration), arXiv:1409.2659 [nucl-ex].
%
\bibitem{WASA2015}
P. Adlarson {\it et al.} (WASA-at-COSY Collaboration), Phys. Rev. C {\bf 91}, 015201 (2015).
%
\bibitem{WASA2006}
M. Bashkanov  {\it et al.}, Phys. Lett. B {\bf 637}, 223 (2006).
%
\bibitem{WASA2012}
P. Adlarson {\it et al.} (WASA-at-COSY Collaboration), Phys. Rev. C {\bf 86}, 032201 (2012).
%
\bibitem{WASA2009}
S. Keleta {\it et al.} (WASA-at-COSY Collaboration), Nucl. Phys. A {\bf 825}, 71 (2009).
%
\bibitem{WASA2014}
P. Adlarson {\it et al.} (WASA-at-COSY Collaboration \&  SAID DAC), Phys. Rev. Lett. {\bf 112}, 202301 (2014).
%
\bibitem{WASA2014-2}
P. Adlarson {\it et al.} (WASA-at-COSY Collaboration \&  SAID DAC), Phys. Rev. C {\bf 90}, 035204 (2014).
%
\bibitem{Dyson64}
F. J. Dyson and N. H. Xuong, Phys. Rev. Lett. {\bf 13}, 815 (1964).
%
\bibitem{YZYS99}
X. Q. Yuan, Z. Y. Zhang, Y. W. Yu, and P. N. Shen, Phys. Rev. C {\bf 60}, 045203 (1999); L. R. Dai, Chin. Phys. Lett. {\bf 22}, 2204 (2005). 
%
\bibitem{Gal2013}
A. Gal and H. Garcilazo, Phys. Rev. Lett. {\bf 111}, 172301 (2013).
%
\bibitem{Gal2014}
A. Gal and H. Garcilazo, Nucl. Phys. A {\bf 928}, 73 (2014).
%
\bibitem{HPW2014}
H. X. Huang, J. L. Ping, and F. Wang, Phys. Rev. C {\bf 89}, 034001 (2014).
%
\bibitem{CHX2015}
H. X. Chen, E. L. Cui, W. Chen, T. G. Steele, and S. L. Zhu, Phys. Rev. C {\bf 91}, 025204 (2015).
%
\bibitem{DZYW2003}
L. R. Dai, Z. Y. Zhang, Y. W. Yu, and P. Wang, Nucl. Phys. A {\bf 727}, 321 (2003).
%
\bibitem{HZY2006}
F. Huang, Z. Y. Zhang, and Y. W. Yu, Phys. Rev. C {\bf 73}, 025207 (2006).
%
\bibitem{ZYS97}
Z. Y. Zhang, Y. W. Yu, P. N. Shen,  L. R. Dai, A. Faessler, and U. Straub, Nucl. Phys. A {\bf 625}, 59 (1997).
%
\bibitem{amk91}
A. M. Kusainov, V. G. Neudatchin, and I. T. Obukhovsky, Phys. Rev. C {\bf 44}, 2343 (1991).
%
\bibitem{abu91}
A. Buchmann, E. Fernandez, and K. Yazaki, Phys. Lett. B {\bf 269}, 35 (1991).
%
\bibitem{emh91}
E. M. Henley and G. A. Miller, Phys. Lett. B {\bf 251}, 453 (1991).
%
\bibitem{HZY2004}
F. Huang, Z. Y. Zhang, and Y. W. Yu, Phys. Rev. C {\bf 70}, 044004 (2004).
%
\bibitem{HZ2004}
F. Huang and Z. Y. Zhang, Phys. Rev. C {\bf 70}, 064004 (2004).
%
\bibitem{Huang2014}
F. Huang, P. N. Shen, Z. Y. Zhang, and W. L. Wang, arXiv:1408.0458.
%
\bibitem{LSZY2001}
Q. B. Li, P. N. Shen, Z. Y. Zhang, and Y. W. Yu, Nucl. Phys. A {\bf 683}, 487 (2001).
%
\bibitem{OY80}
M. Oka and K. Yazaki, Phys. Lett. B {\bf 90}, 41 (1980).
%
\bibitem{Kusainov91}
A. M. Kusainov {\it et al.}, Phys. Rev. C {\bf 44}, 2343 (1991).
%
\bibitem{Glozman93}
L. Ya. Glozman, V. G. Neudatchin, and I. T. Obukhovsky, Phys. Rev. C {\bf 48}, 389 (1993).
%
\bibitem{Stancu97}
Fl. Stancu, S. Pepin, and L. Ya. Glozman, Phys. Rev. C {\bf 56}, 2779 (1997).
%
\bibitem{Dong2015}
Y. B. Dong, P. N. Shen, F. Huang, and Z. Y. Zhang, arXiv:1503.02456.
%
\bibitem{BBC2013}
M. Bashkanov, S.J. Brodsky, and H. Clement, Phys. Lett. B {\bf 727}, 438 (2013).
%
\bibitem{Itonaga1996}
K. Itonaga, A. J. Buchmann, G. Wagner, and A. Faessler, Nucl. Phys. A {\bf 609}, 422 (1996).
%
\bibitem{Obukhovsky1997}
I. T. Obukhovsky, K. Itonaga, G. Wagner, A. J. Buchmann, and A. Faessler, Phys. Rev. C {\bf 56}, 3295 (1997).
%
\bibitem{BCS2015}
M. Bashkanov, H. Clement, T. Skorodko, arXiv:1502.07156.
%
\bibitem{BCS2015-2}
M. Bashkanov, H. Clement, T. Skorodko, arXiv:1502.07500, and references therein.
%
\end{thebibliography}
\end{document}